\pdfoutput=1
\documentclass[twocolumn,nofootinbib,prl]{revtex4}

\usepackage{graphicx}
\usepackage{amssymb,amsmath,amsfonts}
\usepackage{bbm,cancel}
\usepackage{hyperref}

\def\beq{\begin{eqnarray}}
\def\eeq{\end{eqnarray}}

\def\Mp{}
\def\MMp{}

\def\rg{{\sqrt{\scalebox{0.75}[1.0]{\( - \)}|g|}}}

\newcommand{\sect}[1]{{\em {#1.}}}

\begin{document}

\title{Small dark energy without small parameters}
\author{Benjamin Shlaer}
\email{shlaer@cosmos.phy.tufts.edu}
\affiliation{Department of Physics, The University of Auckland, Private Bag 92019, Auckland, New Zealand}
\affiliation{Institute of Cosmology, Department of Physics and Astronomy,\\Tufts University, Medford, MA  02155, USA}
\date{\today}

\begin{abstract}\noindent
We present a prototype model that resolves the cosmological constant problem using matter alone, i.e., without modifying gravity.  
Its generic cosmological solutions
 adjust an arbitrarily large, negative dark energy to a positive value parametrically suppressed by an initial field velocity.  
 Inflationary initial conditions lead to a positive dark energy exponentially smaller in magnitude than any model parameter, 
 or any scale in the initial conditions.  
\end{abstract}


\maketitle

\sect{Introduction}
The observed \cite{Riess:1998cb, Perlmutter:1998np} value of the dark energy density is problematic for two reasons.  
First, known contributions to the vacuum energy density are so large in magnitude that the implied delicate
cancellations between them appear highly improbable  (cosmological constant problem  \cite{Weinberg:1988cp, Martin:2012bt,Burgess:2013ara}).  
Second, even if such a cancellation could be orchestrated via a 
sufficiently rich landscape of solutions, cosmological initial conditions are exceedingly unlikely to correctly navigate this landscape (coincidence problem \cite{Peebles:2002gy, Weinberg:2000yb}).
In brief, the cosmological constant problem is about the particle physics of empty space, and the coincidence problem is about the dynamics of an expanding universe containing 
a {\em dynamical} dark energy, meaning dark energy with a non-trivial discrete or continuous configuration space.

The cosmological constant problem is best described in the language of effective field theory, where every energy scale $\mu$ in particle physics contributes a term of order $\pm\mu^4$ to
the vacuum energy density.  The electroweak scale contributes a term nearly 60 orders of magnitude larger than the observed value, 
which implies there must be a tremendously lucky cancellation occurring without a mechanism.  Most troubling, new physics at higher energy scales makes
the tuning problem worse \cite{Burgess:2013ara}.

A direct attempt at a solution involves ``relaxing'' a large positive cosmological constant dynamically \cite{Dolgov:1982qq,Ford:1987de, Dolgov:1996zg, Charmousis:2011bf}.
Unfortunately, this program has only succeeded at producing flat space by sending Newton's constant to zero.  
Famously, flat space is not an equilibrium solution to any  natural  scalar field theory coupled to gravity \cite{Weinberg:1988cp}.   
Partial relaxation mechanisms \cite{Tsamis:1992sx,Abramo:1997hu} do not predict a small cosmological constant.
More generally, a seemingly necessary feature of any model solving the cosmological constant problem is a tremendously 
dense landscape of values the dark energy density can take 
\cite{Abbott:1984qf, Anderson:1971pn, vanderBij:1981ym, Henneaux:1989zc, Kaloper:2015jra, Alberte:2016izw}.  
Then the problem becomes how initial conditions can be expected to produce the special values \cite{Wetterich:1987fm, Peebles:1987ek, Zlatev:1998tr,Weinberg:2000yb}.

Assuming
the vacuum energy could be approximately canceled by some other dynamical energy density, e.g. Abbot's linear washboard 
potential \cite{Abbott:1984qf}, cosmological initial conditions would have to be exceedingly fine-tuned in order for standard big-bang 
cosmology to successfully hide the effects of dark energy until the universe was extremely large by early-universe standards.
This is because the energy density in particles dilutes much more rapidly than dark energy during cosmological expansion.  
Generic initial conditions lead to dark energy domination far too soon, after an expansion factor of only a few.   
If negative, dark energy would lead to a collapsing universe.  If positive, it leads to accelerated expansion of the universe.

The most developed solution to the coincidence problem is based on the anthropic principle \cite{Carter:1974zz}, meaning the apparent fine-tunings 
in the initial conditions of the visible universe are the result of an environmental (selection) effect: 
Observers might only exist in the vanishingly small subset of configuration space where dark energy is small.    
 This approach \cite{Weinberg:1987dv,Vilenkin:1994ua} was buoyed by the confirmation of its apparent prediction of a
 non-zero dark energy \cite{Riess:1998cb, Perlmutter:1998np}.  
Producing at least one habitable region within a primarily hostile theory could be accomplished via eternal inflation \cite{Vilenkin:1983xq, Linde:1986fc}, although predictivity itself must
first be resuscitated by addressing the measure problem \cite{Linde:1993xx, Guth:2000ka}.

If one finds the anthropic principle premature or the measure problem too formidable, then standard cosmology, i.e., the null energy condition (NEC), 
must be abandoned \cite{Steinhardt:2006bf,Biswas:2009fv,Alberte:2016izw}.
Otherwise, there is no natural way around the coincidence problem, due to the following argument.
The only probe of vacuum energy is its gravitational effect, and so dynamical adjustment should take place whenever dark energy has become a significant fraction
of the total energy density, causing strong accelerated expansion.  Hence, soon after premature acceleration begins, the dark energy density needs to decrease \cite{Dodelson:2001fq}.  
However the adjustment mechanism must no longer work today, since the observed dark energy density does not appear to be decreasing.  
Thus the unnaturally small value of today's dark energy must be a model parameter used to shut off the mechanism \cite{Alberte:2016izw}.

In this Letter, we show with a specific model how both the cosmological constant and coincidence problems can be solved {\em without introducing small parameters.\/}  
The model solves the cosmological constant problem by having a continuum landscape \cite{Shlaer:2014gna}, just like unimodular gravity \cite{Anderson:1971pn, vanderBij:1981ym, Henneaux:1989zc}. 
The model solves the coincidence problem as well,  evading the difficulty mentioned in the previous paragraph by 
adjusting an initially large but {\em negative} dark energy via a nonstandard temporarily-bouncing cosmology \cite{Steinhardt:2006bf}.  
This provides a natural shut-off for the adjustment mechanism, namely dark energy becoming positive \cite{Biswas:2009fv}.
Finally, the model does not need fine-tuned parameters or initial conditions because the hierarchy between initial conditions and today's dark energy density is set by
an initial field velocity, which is naturally set to an exponentially small value via cosmological inflation.

\sect{The model}
Our self-tuning model (introduced in Ref.~\cite{Shlaer:2014gna} as a solution to the problem of time) 
consists of Einstein gravity minimally coupled to two scalar fields, $\Lambda$ and $\chi$ via
\beq
S = \int\rg{\rm d}^4x\left( \frac{\MMp R}{2} - \partial_\mu\Lambda \partial^\mu\chi  - \Lambda \right) + S_{\rm mat}.
\eeq
We work in reduced Planck units, where $8\pi G = 1$.  
The equations of motion are
\beq
\nabla_\mu\nabla^\mu \Lambda &=& 0,\\
\nabla_\mu\nabla^\mu \chi &=& 1,\\
\MMp G_{\mu\nu} &=& -\Lambda g_{\mu\nu} + T_{\mu\nu}^{\rm nec} + T_{\mu\nu}^{\rm mat},
\eeq
where the NEC violating \cite{Rubakov:2014jja} stress tensor component is
\beq
T_{\mu\nu}^{\rm nec} = 2\partial_{(\mu}\Lambda\,\partial_{\nu)}\chi  -  g_{\mu\nu}\partial^\sigma\Lambda\partial_\sigma\chi.
\eeq
Note that any constant $\Lambda$ is a solution.  
Because any renormalization of the vacuum energy can just be absorbed
into a field redefinition $\Lambda\;\mapsto\;\Lambda+{\rm const.}$, there is no cosmological constant problem.  (We do not assume $\Lambda$ is initially small.)

Note that there is a ghost instability \cite{Hsu:2004vr, Cline:2003gs, Carroll:2003st} due to the
wrong-sign \cite{Feng:2004ad} canonical kinetic term for the one of the two scalar linear combinations $\phi_-=(\Lambda-\chi)/\sqrt{2}$.  
Classically, this is harmless, and
it can be shown that on an arbitrary initial Cauchy surface, if both $\phi_-$ and $\dot\phi_-$ are chosen to be spatially
constant, their solution remains spatially constant in geodesic slicing for arbitrary geometry. 
Hence, restricting to such initial conditions guarantees classical stability.  However, quantum fluctuations of $\phi_-$ about any classical solution must
be canceled in order for the theory to be viable.  This is accomplished with the addition of  Becchi-Rouet-Stora-Tyutin (BRST) ghosts, 
as proposed in Ref.~\cite{Nojiri:2016mlb} and developed in Ref.~\cite{Mori:2017dhe}, although we suppress the anti-commuting fields here.
Alternative possible cures for the ghost instability include a Lagrangian with higher powers of the kinetic term \cite{ArkaniHamed:2003uy},
or higher derivative kinetic terms \cite{Nicolis:2008in, Deffayet:2009wt, Ijjas:2016tpn, Creminelli:2016zwa}, which we leave for future work.

Restricting to homogeneous isotropic cosmologies, the metric is described by the FLRW ansatz
\begin{align}
{\rm d s}^2 = -{\rm d}t^2 + a^2(t)\left(\frac{{\rm d}r^2}{1- k r^2} + r^2 {\rm d}\Omega_2^2\right).
\end{align}
Overdots denote derivatives with respect to cosmic time $t$.
The scalar equations of motion can be integrated once to yield
\begin{align}\label{eq:lambdadot}
\dot \Lambda = -p/a^3(t),  \qquad \dot \chi = -V(t)/a^3,
\end{align}
where $p$ is an arbitrary integration constant, and 
\begin{align}
V(t) = \int a^3 {\rm d} t
\end{align}
 is proportional to the four-volume of the universe at time $t$, plus an integration constant.

If we choose $S_{\rm mat.}$ to describe a perfect fluid with equation of state $w = -1-\frac{\partial\log\rho_{\rm mat}}{\partial\log a^3}$, the constraint equation is
\begin{align}\label{eq:constraint}
H^2= -\frac{k}{a^2} + \frac{1}{3\MMp}\left[\frac{p V(t)}{a^6} + \Lambda(t) + \frac{\rho_{\rm mat}^0}{a^{3(1+w)}}  \right],
\end{align}
where $H = \dot a/a$ and $\rho_{\rm mat}^0$ is the matter density when $a=1$.  
The second-order Friedmann equation is
\begin{align}
\frac{\ddot a}{a} =& -\frac{1}{6\MMp}\left[ \frac{4 p V(t)}{a^6} - 2 \Lambda(t)  + (1+3w)\frac{\rho_{\rm mat}^0}{a^{3(1+w)}} \right] ,
\end{align}
from which it can be deduced that there are two interacting fluids associated with the scalar fields.  Their energy densities are $pV a^{-6}$ with equation of state $w_{pV}=1$, and $\Lambda$ with equation of state $w_{\Lambda}=-1$.  We will refer to the former as the $pV$ fluid.  Note that the effective equation of state of the $pV$ fluid is less than one during expansion and larger than one during contraction due to the increasing four volume $V(t)$.

\sect{Cosmological evolution}
We will initially focus on the cases where matter is dust  ($w=0$) or radiation ($w=1/3$), and $k=0$.
The interesting cosmological solutions are those whose initial conditions have  
$\Lambda$ negative but with a positive time derivative.
In this case, both the $pV$ fluid 
and the $\Lambda$ fluid have negative energy, while ordinary matter $\rho_{\rm mat}$ has positive energy.  After sufficient expansion, $\Lambda$ will eventually
dominate, leading to turnaround and a contracting phase.  Since the $pV$ fluid has a stiff equation of state and negative energy density, it ends the contracting phase with a smooth bounce.  Hence, the universe
temporarily resembles a cyclic model \cite{Steinhardt:2001st}.  Since $\Lambda$ is increasing with time, each subsequent turnaround occurs at larger and larger scale factor, leading to a significant increase
in four volume $V$, and as a consequence, in the magnitude of the $pV$ fluid.  This in turn causes the bounce to occur at larger and larger scale factor, slowing the increase in $\Lambda$,
as seen in Eq.~(\ref{eq:lambdadot}).
The process repeats with $\Lambda$ ever more slowly approaching zero until the final bounce that sends it positive.  Because the scale factor then grows
rapidly, $\Lambda$ quickly approaches its asymptotic value, which we call $\Lambda_f$.

The two\footnote{In our convention there is no large {\em initial\/} hierarchy between $\rho_{\rm mat.}$ and $\Lambda$.} important initial conditions are the initial field velocity $\dot\Lambda_i$ and the initial hierarchy between $\Lambda$ and the $pV$ fluid energy density.  
The former determines how small today's dark energy $\Lambda_f$ will be, and the latter determines the initial amplitude of oscillation in the scale factor.

\begin{figure}[tbp]
   \centering
   \includegraphics[width=3.4in]{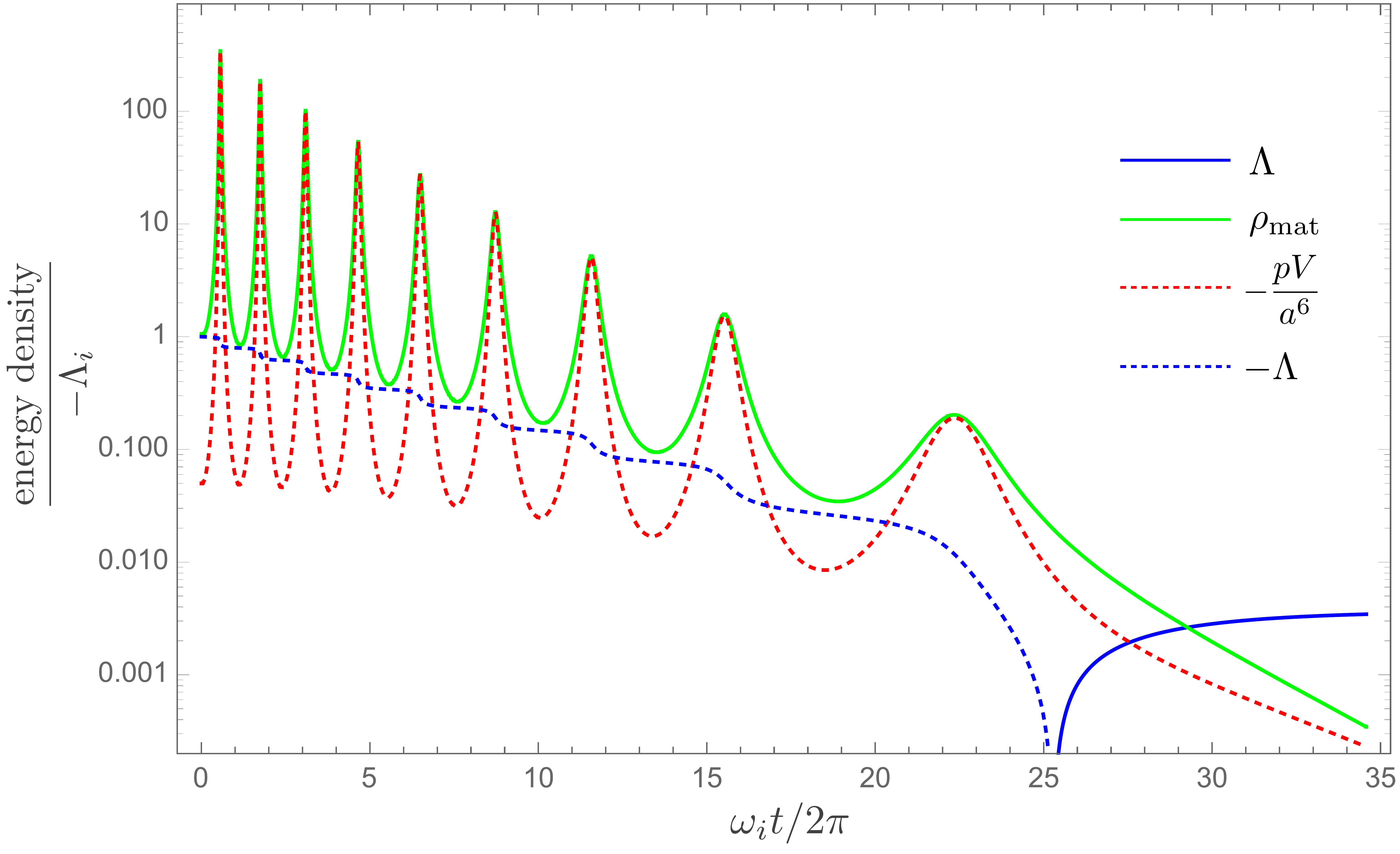} 
   \caption{Numerical evolution of component energy densities for perfect fluid $\rho_{\rm mat}$ with $w = 1/3$.  
   The final dark energy density is suppressed relative to its initial value $\Lambda_i$ by the initial field velocity $\dot \Lambda_i$.}
   \label{fig:energydensities}
\end{figure}

Let us define the frequency 
\begin{align}
\omega = (1+w)\sqrt{-3\Lambda},
\end{align}
which is equal to the scale factor oscillation frequency when its amplitude is large.  Then the final value of dark energy is given by
 \begin{align}
 \Lambda_f&\sim  -\left(\frac{\dot\Lambda_i}{-\Lambda_i\omega_i}\right)^{2}  \Lambda_i\quad &{\rm for}\;w = 0,\label{eq:matterhierarchy}\\
  \Lambda_f&\sim  -\left(\frac{\dot\Lambda_i}{-\Lambda_i\omega_i}\right)^{4/3}  \Lambda_i\quad &{\rm for}\;w = 1/3.\label{eq:radhierarchy}
 \end{align}
At this point, we have parametric control over the initial-to-final dark energy hierarchy in terms of the initial field velocity $\dot \Lambda_i$. 
We plot a numerical example in Fig.~\ref{fig:energydensities}.

\sect{Solutions and dynamics}
When the scale factor oscillates with a large amplitude, we can safely neglect one of the three fluids at any given time.
This lets us write down approximations for the increase in $\Lambda$, $V$, and $t$ per cycle as 
\begin{align}
 \delta\Lambda \sim \Mp\sqrt{\frac{-p}{V}},  \quad \delta V \sim \Mp\frac{\left(\frac{\rho_{\rm mat}^0}{-\Lambda}\right)^{\tfrac{1}{1+w}}}{\sqrt{-\Lambda}}, \quad \delta t \approx\frac{2\pi}{\omega} .
\end{align}
 From the above equations it is clear that $\Lambda$ increases in smaller steps as $V$ gets large, and that $V$ increases in larger steps as $-\Lambda$ gets small.

There is a non-oscillating exact solution for pressureless matter $(w=0)$ and zero spatial curvature given by
\begin{align}
a(t)&= a_i\exp\left(\sqrt{\frac{\Lambda_f}{3}}\,t\right),\quad
\rho_{\rm mat}(t)= \frac{2a_i^3\left(\Lambda_f-\Lambda_i\right)}{a^3(t)},\\
\Lambda(t)&= -\frac{\rho_{\rm mat}(t)}{2} + \Lambda_f, \qquad\,\,
\frac{pV(t)}{a^6(t)}=-\frac{\rho_{\rm mat}(t)}{2}.
\end{align}
This lets us verify that Eq.~(\ref{eq:matterhierarchy}) is exact to lowest order in $\Lambda_f/\Lambda_i$.  Large amplitude oscillations
only modify it by a factor of order unity.

Another exact solution exists for $\rho_{\rm mat}=0$ and negative spatial curvature. (Or equivalently, for matter with ${w = -1/3}$.)
In this case, $\Lambda \to 0$ as $t \to \infty$:
\begin{align}
a(t) &= t/t_0\\
k &=  p t_0/4\MMp - 1/t_0^2\\
\Lambda(t) &=  p t_0^3/2 t^2\\
V(t) &= t^4/4 t_0^3.
\end{align}
There is a big bang singularity at $t = 0$, where all three fluids have divergent energy density.
The energy density in the $pV$ component is always half the energy density in the $\Lambda$ component, and the spatial curvature ``energy density''
is related to the $\Lambda$ energy density by a constant factor,
\begin{align}
\rho_k = -\frac{3 \MMp k}{a^2} = -\left(\frac{3}{2} - \frac{6\MMp}{p t_0^3}\right)\Lambda.  
\end{align}
Note that this (unrealistic) scale invariant cosmology has neither a flatness problem nor a cosmological constant problem:  All three fluids are always related by time-independent factors of order unity.
The asymptotic solution is flat space, and Weinberg's theorem~\cite{Weinberg:1988cp} is evaded since $\chi$ does not have an equilibrium solution.

For the case ${\Lambda_i < 0}$, both of the above exact solutions are attractors in the following sense:  
Any perturbation of the initial conditions causes the scale factor to oscillate about the above exact solutions, with a constant
amplitude.  Because the mean value of the scale factor is growing, the constant amplitude oscillations in $a(t)$ represent 
decaying oscillations in $\log\left(a(t)\right)$, as visible in Fig.~\ref{fig:energydensities}.  Note that as long as negative spatial curvature is
dominant, $\Lambda$ will approach zero but never become positive.  However, matter can be produced at a later
time via cosmological hysteresis \cite{Kanekar:2001qd}, and this can send $\Lambda$ positive.   Note that such a mechanism
is also able to solve the flatness problem, although it might be more economical to use inflation \cite{Guth:1980zm}.

\sect{Inflationary suppression of today's dark energy}
Because the field velocity $\dot \Lambda$ scales like $a^{-3}$, any mechanism which leads to large expansion will, according to Eqs.~(\ref{eq:matterhierarchy})-(\ref{eq:radhierarchy}),
lead to a small final dark energy.  Inflation does just this.
Inflationary initial conditions are described by a temporary phase when the matter equation of state is roughly ${w\approx-1}$.  
The constraint Eq.~(\ref{eq:constraint}) implies the initial matter density is large enough to
compensate the assumed large negative initial $\Lambda$, meaning $0\;<\;{\rho_{\rm inf}=\left(\rho_{\rm mat}+\Lambda\right)_{i}}$.  
After the scale factor has increased by a factor of $\exp(N)$, reheating occurs, whereby the matter equation of state evolves to ${w=1/3}$.
Provided the initial field velocity satisfies
\begin{align}\label{eq:Lambdadotbound}
0< \dot \Lambda_{i} \lesssim -H_{\rm inf} \Lambda_{i},
\end{align}
where $H_{\rm inf} = \sqrt{\rho_{\rm inf}/3}$,
inflation ends with a negative cosmological constant which is of order $\Lambda_i$.  Since after $N$ e-folds of inflation the field velocity redshifts to
\begin{align}
\dot \Lambda_{\rm reheat} = \exp(-3N)\dot\Lambda_{i}, 
\end{align}
the value for today's dark energy is
\begin{align}
\Lambda_f &\sim -\exp(-4N)\frac{\dot\Lambda_i^{4/3}}{\Lambda_i}.
\end{align}
Thus inflation provides an exponential hierarchy between today's dark energy and the scale of initial conditions.
Furthermore, using the inequality (\ref{eq:Lambdadotbound}), we obtain
\begin{align}
\Lambda_f &\lesssim \frac{(-\Lambda_i)^{1/3}\rho_{\rm inf}^{2/3}}{\exp(4N)}. 
\end{align}
Finally, if we assume no large initial hierarchy between $\rho_{\rm mat}$ and $\Lambda_i$, then $\Lambda_i \sim -\rho_{\rm inf}$ and
\begin{align}
\Lambda_f \lesssim \frac{\rho_{\rm inf}}{\exp(4N)}.
\end{align}
Hence, a natural upper bound on the amount of inflation (or a lower bound on the scale of inflation) is obtained from today's dark energy, meaning observation of
dark energy amounts to an observation of the initial conditions of inflation.
This raises the possibility of observing other inflationary initial conditions that are suppressed by the amount of inflation \cite{Aslanyan:2015pma}.

\begin{figure}[tbp]
   \centering
   \includegraphics[width=3.4in]{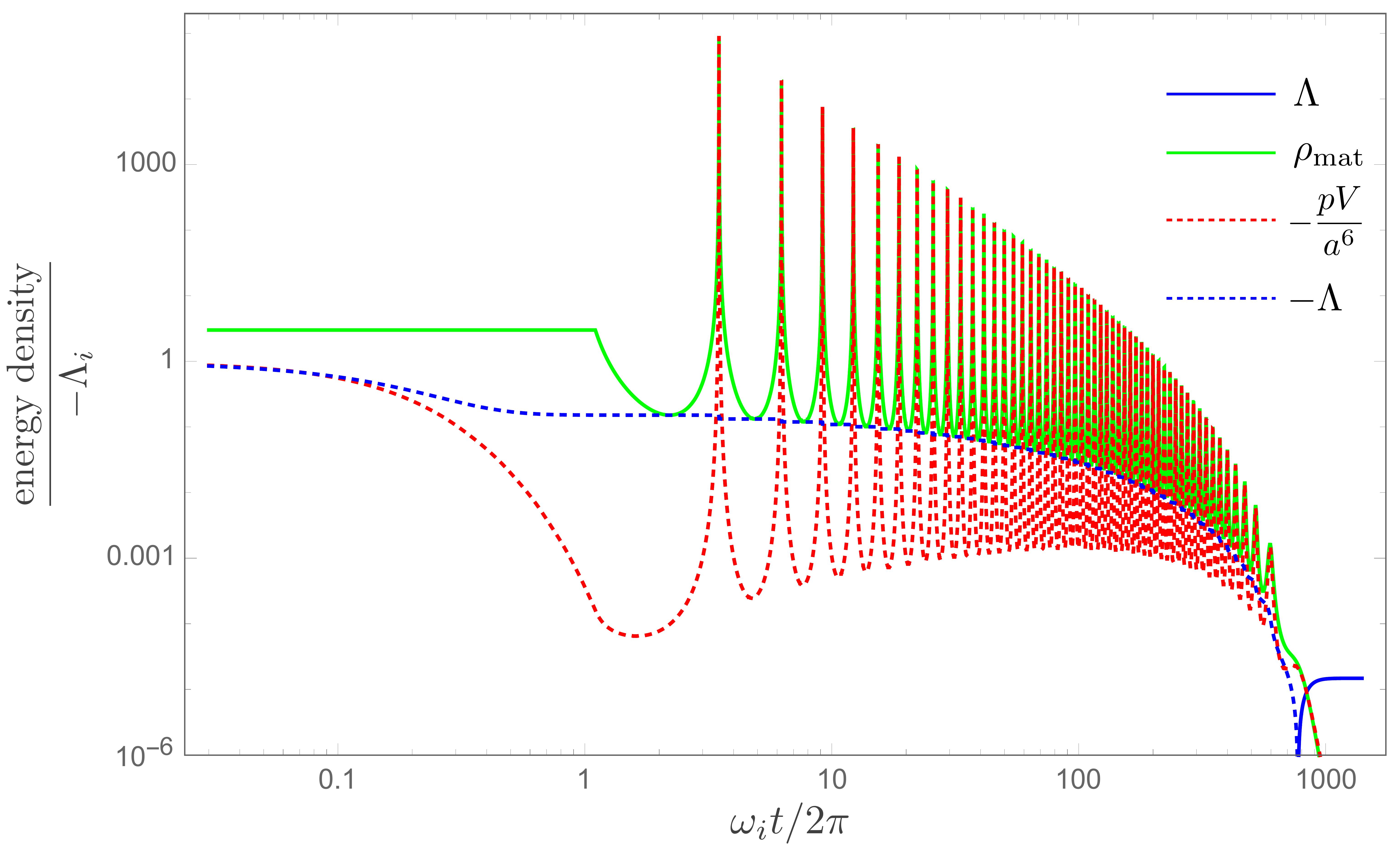} 
   \caption{Numerical evolution of component energy densities for $N=3$ e-folds of inflation with inflationary matter equation of state $w = -1$ reheating abruptly to $w = 1/3$.}
   \label{fig:energydensitiesNe}
\end{figure}

\sect{Challenges}
Bouncing cosmologies face several obstacles \cite{Battefeld:2014uga}.  
Observations rule out a bounce since last scattering, so the minimum acceptable amplitude of the last oscillation is over a thousand.  
Unfortunately despite the fact that inflationary initial conditions
lead to an exponentially large initial bounce amplitude, the model coupled to a perfect fluid $\rho_{\rm mat}$
predicts the amplitude of oscillation of $\log(a)$ to gradually attenuate, becoming only of order unity for the last bounce, as visible in Fig.~\ref{fig:energydensitiesNe}.  
This rules out the model.  A possible fix is to include an additional positive energy field with equation of state ${w\;\sim\;1}$ which can mimic an extreme reduction in 
the magnitude of the $pV$ fluid.  
One possibility is a fuzzy dark matter axion \cite{Hu:2000ke}, since it can experience cosmological hysteresis \cite{Kanekar:2001qd} from bounces in the late 
universe due to its very small mass.  
Since this increases its energy density, it also deepens the bounces, which encourages $\Lambda$ to become positive at a time set by dark matter length scales.

The model as described (or improved with sufficiently deep bounces) must additionally be made compatible with big bang nucleosynthesis and 
evolution of density perturbations \cite{Brandenberger:2009ic}, neither of which we attempt here.
Note that we can avoid having to merging the bouncing scenario presented here with standard cosmology simply by following the example of Ref.~\cite{Alberte:2016izw},
where a NEC violating mechanism restarts inflation after the cosmological constant has been correctly set. 
The lesson from this model is that tuning dark energy from large negative to small positive values is apparently easier than
tuning it from large positive to small positive values.

Although large amplitude recent bounces can be compatible with observational cosmology, another possibility
is to seek a more conventional cosmology with an alternative two-scalar Lagrangian with the following properties:
\begin{enumerate}
\item A linear potential $U(\Lambda)=\Lambda$ to absorb renormalization of the vacuum energy.
\item An equation of motion $\frac{\delta S}{\delta\chi}=0$ which allows any constant $\Lambda$ to be a solution.
\item Sequestering of ghost and gradient instabilities \cite{Mori:2017dhe}.
\end{enumerate}
The second property is achieved by having all $\chi$-dependent terms in the Lagrangian be proportional to $\partial_\mu\Lambda$,
so they vanish when $\Lambda$ is a constant.
Brief moments of dark energy domination can be compatible with CMB constraints \cite{Pettorino:2013ia}, 
so the hope would be that the tuning of $\Lambda$ would occur not from deep bounces, but from repeated sudden decreases in $H^2$,
which occurs whenever negative dark energy begins to dominate.  Such behavior could arise from scalar-tensor kinetic terms like $G^{\mu\nu}\partial_\mu\Lambda\partial_\nu\chi$. 

We have demonstrated that a very simple model can solve both the cosmological constant and coincidence problems.  The challenge of finding a realistic cosmology within our framework remains.  Two
of the most promising scenarios, in our opinion, are to have extremely deep bounces, or to avoid bounces altogether via scalar-tensor couplings to $\Lambda$.

\acknowledgments
We thank Jose Blanco-Pillado, Robert Brandenberger, Richard Easther, Mark Hertzberg, and Alex Vilenkin for helpful discussions.
This work was supported by a grant from the Foundational Questions Institute.

\bibliography{smalldarkenergy}



\end{document}